
------------------------------------------------------------------------------
\magnification=\magstep1
\nopagenumbers
\tolerance=1000
\hyphenpenalty=3000
\hsize 15truecm
\vsize 22truecm
\parskip 5pt plus 2pt minus 2pt
\font\twelverm=cmr10 scaled\magstep3
\font\twelvebf=cmbx10 scaled\magstep3
\font\twelveit=cmti10 scaled\magstep3
\font\twelvesl=cmsl10 scaled\magstep3
\font\twelvemus=cmmi10 scaled\magstep3
\font\grandrm=cmr10 scaled\magstep1
\font\grandbf=cmbx10 scaled\magstep1
\font\grandit=cmti10 scaled\magstep1
\font\grandsl=cmsl10 scaled\magstep1
\font\grandmus=cmmi10 scaled\magstep1
\font\tenrm=cmr10 scaled\magstep0
\font\tenbf=cmbx10 scaled\magstep0
\font\tenit=cmti10 scaled\magstep0
\font\tensl=cmsl10 scaled\magstep0
\font\tenmus=cmmi10 scaled\magstep0
\font\eightrm=cmr8
\font\eightbf=cmbx8
\font\eightit=cmti8
\font\eightsl=cmsl8
\font\eightmus=cmmi8
\def\bigtype{\let\rm=\twelverm \let\bf=\twelvebf
 \let\it=\twelveit \let\sl=\twelvesl \let\mus=\twelvemus}
\def\grande{\let\rm=\grandrm \let\bf=\grandbf
 \let\it=\grandit \let\sl=\grandsl \let\mus=\grandmus}
\def\medtype{\let\rm=\tenrm \let\bf=\tenbf
 \let\it=\tenit \let\sl=\tensl \let\mus=\tenmus}
\def\smalltype{\let\rm=\eightrm \let\bf=\eightbf
 \let\it=\eightit \let\sl=\eightsl \let\mus=\eightmus}

\footline = {\hss\bf \folio\hss}
\def\ref{\par\noindent\hangindent 20pt}

\def\pasp{{\it Pub. Astr. Soc. Pac.}\ }

\def\>{\hskip 0.6em}

\def\lsim{\, \lower2truept\hbox{${<
\atop\hbox{\raise4truept\hbox{$\sim$}}}$}\,}
\def\gsim{\, \lower2truept\hbox{${>
\atop\hbox{\raise4truept\hbox{$\sim$}}}$}\,}
\baselineskip=12pt
\rm
\centerline { }
\vskip 1truein
\centerline  {\bf\bigtype SKY SUBTRACTION WITH FIBER-SPECTROGRAPHS
\medtype\footnote{*}
{\rm Based on observations collected at the European Southern
Observatory, La Silla, Chile.}}
\vskip 1truecm
\centerline{\bf Claudio LISSANDRINI}
\centerline{Dipartimento di Astronomia della Universit\`a di Padova}
\centerline{Vicolo dell' Osservatorio 5, I-35122 Padova, Italy.}
\centerline{\it email: lissandrini@astrpd.astro.it }
\vskip 1truecm
\centerline{\bf Stefano CRISTIANI }
\centerline {Dipartimento di Astronomia della Universit\`a di Padova}
\centerline {Vicolo dell' Osservatorio 5, I-35122 Padova, Italy.}
\centerline {\it email: cristiani@astrpd.astro.it }
\vskip 1truecm
\centerline{\bf Fabio LA FRANCA }
\centerline {European Southern Observatory}
\centerline {K.Schwarzschild-Str.2, D-85748 Garching bei M\"unchen, FRG.}
\centerline {\it and}
\centerline {Dipartimento di Astronomia della Universit\`a di Padova}
\centerline {Vicolo dell' Osservatorio 5, I-35122 Padova, Italy.}
\centerline {\it email: lafranca@astrpd.astro.it }
\vskip 3mm
\vskip 1cm
\centerline { August 1994}
\vskip 7mm
\centerline {To appear in the \pasp}
\vskip 7mm
\centerline {Send offprint request and proofs to S.Cristiani}
\vskip 7mm
\baselineskip=16pt
\vfill\eject

{\bf SUMMARY}

The sky subtraction performances of multi-fiber spectrographs are discussed,
analysing in detail the case of the OPTOPUS system at the 3.6 meter ESO
telescope at La Silla.
A standard technique, based on flat-fields obtained with a
uniformly illuminated screen on the dome, provides poor results.
A new method has been developed, using the [OI] emission line at 5577 \AA\
as a calibrator of the fiber transmittance,
taking into account the diffuse light
and the influence of each fiber on the adjacent ones, and
correcting for the effects of the image distortions on the sky sampling.
In this way the accuracy of the sky subtraction
improves from $2-8 \%$ to $1.3-1.6 \%$.

\medskip

{\bf 1. INTRODUCTION}

Spectrographs fed by optical fibers allow to observe many (from tens to
hundreds) objects simultaneously.
The higher the spatial density of the targets, the larger the advantage
over ``classic'' single-slit spectroscopy, thus
a considerable effort has been spent in extending fiber spectroscopy
to fainter and fainter objects. To this end and in view
of the application of this technique to telescopes of the eight-ten meter
class, the limitations imposed by the accuracy in the sky
subtraction have become a key issue.

Various papers (Elston and Barden 1989, Wyse and Gilmore
1992, Mignoli and Cuby 1994) have discussed this point with the following
results:

\item {1)} An accuracy of the sky subtraction in the range $1-2\%$ is
considered good in practice, while, on theoretical grounds, a limiting accuracy
of a percent or better is expected to be achievable.

\item {2)} Such a limitation is due to:  focal ratio degradation of the fibers,
internal scattered light, sky variations across the field
(caused by blue galaxies, airglow, etc.), telecentricity
\medtype\footnote{$^{\dag}$} {\rm {\it Telecentricity}: effect due to the
misalignment between the axis of the optical fibers and the light beam axis
coming from the telescope output pupil.} effects (Wynne 1993),
contamination from adjacent fibers (cross-talk), poor determination of the
fiber
transmittance and of the wavelength calibration.

Section 2 describes a ``standard'' approach to the sky
subtraction and how to measure the accuracy of the method.
In section 3 the importance of the various sources of noise is analyzed.
Section 4 outlines the basics of a new sky subtraction method and
presents the comparison with the results of section 3.
Finally, section 5 deals with further problems regarding the extraction of
the spectra.
\medskip

 {\bf 2. ESTABLISHING THE ACCURACY OF THE SKY SUBTRACTION}

{\it 2.1) the instrument}

In the following we will analyze the sky subtraction performances of a
multifiber spectrograph in the case of the OPTOPUS system operating at the
3.6 meter telescope of ESO La Silla, described by Avila and D'Odorico (1993)
and in the ESO Users Manual (Schwarz and Melnick, 1993).
The instrument consists of 50 fibers, manually plugged in a previously
drilled metal template and feeding a B. \& C. grating spectrograph.
Each fiber has a diameter of 320 $\mu m$, corresponding to 2.3 arcsec on
the sky.
A typical raw CCD \footnote{$^{\dag}$}
{In the present observations a CCD ESO TEK $571\times 520$, RON=$8.8 e^-$,
GAIN=$1.7 e^-/ADU$, pixels of $27 \mu m$ was used.}
frame obtained with OPTOPUS with a scale of 9 \AA /pixel (resolution of 21 \AA)
is shown in Figure 1.

\vskip 1cm
\centerline{Figure 1}
\vskip 1cm

There is a  fundamental difference between long-slit and multifiber
spectrographs: in the former, the sky spectrum is sampled on both sides of
the target; in the latter some fibers are dedicated to the various targets,
while the sky is sampled by a suitable number of different fibers, each
with its own transmittance. In a multifiber spectrograph, in addition
to the standard operations applied for the reductions of long-slit data,
to evaluate the ``true sky'' it is necessary to determine the transmittance
of each fiber and the influence of each fiber on the adjacent ones
(cross-talk).

Together with the ``astronomical'' frames, a few sets of
calibration frames are obtained: bias and dark exposures, dome flat-fields,
internal-lamp flat-fields and wavelength calibration exposures.
After the bias and dark correction of the frames, individual spectra are
extracted integrating the flux along the columns of each channel
\medtype\footnote{$^*$}
{\rm The term {\it channel} in the following will indicate the sub-image of the
CCD frame containing the signal coming from a given fiber.}
($7$ pixels wide) centered by gaussian fitting of the
transversal profile.
For each frame this produces as many mono-dimensional spectra
as the fibers in use.

{\it 2.2) the ``standard'' method}

Following a ``standard'' approach, in order to determine the relative
transmittance of the fibers we have used
the fluxes recorded in the flat-field spectra, obtained
with diffuse light on the dome (the results obtained with an internal lamp
are definitely worse).

For each sky spectrum we have calculated:

$$ <SKY_i> = <SKYspectrum_i> \cdot {{<\overline{FF}>} \over {<FF_i>}} $$

\noindent where:

\noindent - the symbol $i$ indicates the channel (it identifies the
spectrum and the fiber too);

\noindent - the symbols $<SKYspectrum_i>$ and $<FF_i>$ are
the values of the total flux into the channel between two given pixels
$y_1$ and  $y_2$ along the dispersion axis (e.g. 50 -- 450) for the sky
and the flat-field spectra respectively;

\noindent - the symbol $<\overline{FF}>$ is the average flux of the observed
 flat-fields ${<FF_i>}$;

\noindent - $<SKY_i>$ represents the sky flux estimated from
 the $i^{-th}$ fiber.

We can average over all the $<SKY_i>$, obtaining $<\overline{SKY}>$ and
compute:

$$ (E_{Sky})_i = {{<SKY_i>} \over {<\overline{SKY}>}} - 1 $$

We can adopt the root mean square of $E_{Sky}$ as an estimate of the accuracy
of the sky subtraction.
Such a procedure is implicitly based on a number of hypotheses:

\noindent - the shape of the spectral transmittance $T(\lambda)$ of the fibers
is the same for all the fibers:
$$ T_i(\lambda)/T_j(\lambda) \ne f(\lambda)$$
for each fiber $i$ and $j$;

\noindent - the influence (cross-talk) of each fiber on the adjacent ones is
negligible and there is no stray light into the spectrograph;

\noindent - the variations of the sky due to blue galaxies and airglow
are negligible.

The results obtained for some frames are shown in Table 1.
Two frames for each field were obtained sequentially. The number of fibers
dedicated to the sky varies between 14 and 36.

\vskip 1cm
\centerline{Table 1}
\vskip 1cm

As listed in Table 1 the accuracy of the sky subtraction is greatly
variable ($2-8 \%$) from case to case, showing that the flat-field based
procedures provide typically poor and in any case unreliable results.

{\it 2.3) the ``$5577$ method'' - an improved standard method}

In the spectral range accessible in our observations there are a number of sky
emission lines, the most intense of which is the oxygen line [OI] at
5577.4 \AA.

The [OI] line has the necessary qualifications to be used as an estimator
of the fiber transmittance.
In fact:

\item {1)} its flux is sufficient to reduce the shot noise always below
           $0.6 \%$ and typically at $0.45 \%$ (with a telescope of $3.6$
           meters, a dispersion of 9~\AA /pixel and 40 minutes
           of exposure time);
\item {2)} at the faint magnitudes of interest in the present analysis,
           it is easily measurable both in the object and sky spectra;
\item {3)} it is subject to the same variations of the instrumental
           transmittance as the object flux
           (e.g. differential fiber flexures during the exposure,
that however should not cause variations in the transmittance in a
properly constructed system).

\noindent On the other hand there are also a few drawbacks:

\item {4)} it represents the fiber transmittance only at 5577 \AA;

\item {5)} it occupies only a small area on the CCD and consequently a
``bad'' pixel (a defective pixel, a cosmic-ray hit, etc.)
can lead to a loss of accuracy;
\item {6)} its availability is not guaranteed for all the configurations of
           the spectrograph.

A subtle distinction between the ``5577'' and the ``flat-field'' methods
has to be noted:
while the variance of the latter is due only to errors in the calibration of
the fiber transmittances, the variance of the former is due also to spatial
sky variations. In this way the ``5577'' method potentially provides a more
accurate sky subtraction.

To estimate the accuracy of the sky subtraction we simply replace the
flat-field flux $<FF_i>$ with the [OI] flux $<F5577_i>$, evaluated
by fitting the line profile along the dispersion direction with a suitable
function (a gaussian turns out to be good approximation),
to limit the influence of possible bad pixels.
The results are shown in Table 2.

\vskip 1cm
\centerline{Table 2}
\vskip 1cm

The accuracy of the sky subtraction improves (to $2.1$-$3.0 \%$) except for the
frame C (shown in Figure 1) where an ``anomalous'' fiber is present
(see below); if this
fiber is rejected, the accuracy becomes $3.2 \%$ for this frame too.
Above all the reliability of the results is greatly improved.
\medskip

 {\bf 3. ANALYSIS OF THE RESULTS}

We have investigated various hypotheses to justify the poor results
provided by the ``flat-field'' method:

\item {1)} diffusion and reflections into the spectrograph;

\item {2)} cross-talk among fibers;

\item {3)}  change from fiber to fiber of the shape of the spectral
transmittance  $T(\lambda)$;

\item {4)} saturation and/or non linearity effects of the CCD;

\item {5)} unknown dimensional dependent factors:
possible correlations of the $E_{Sky}$ with the number of the
fiber $i$, with the position of the fibers on the focal plane of
the telescope (right ascension, declination, distance from the
center of the field, ``azimuth'' on the OPTOPUS template);

None of the effects listed in the items $3,4,5$ turned out to be sufficient to
explain the improvement between the ``flat-field'' and the ``5577'' method
(for example Table 3 shows the effect of the variation of the
transmittance as a function of the wavelength range estimated with dome
flat-fields; effects $4$ and $5$ have been measured to be even smaller).

\vskip 1cm
\centerline{Table 3}
\vskip 1cm

\medskip
{\it 3.1 The Sky Concentration}

Let us address the effects of internal diffusion and reflections: we have
grouped these phenomena under the name {\it Sky Concentration} or, in the
following, SC. To extract the SC it is useful to study particular frames
denominated {\it mono-fiber flats} in which only one fiber has been
exposed to the dome or internal-lamp light.
By two dimensional fitting with suitable functions it is possible to evaluate
the SC contribution.
Comparing a number of mono-fiber flats obtained exposing different fibers,
it has been possible to deduce the following properties of the SC:

\item {1)} it is independent of the particular fiber exposed;
\item {2)} it has rotational symmetry;
\item {3)} its contribution is proportional ($2.2\%$) to the total flux in
the frame.

\vskip 1cm
\centerline{Figure 2}
\vskip 1cm

After fitting the SC (Figure 2) with a two dimensional gaussian of
$\sigma_x= \sigma_y = 95 \pm 5~pixels$ the bias-subtracted astronomical
 frame have been corrected by performing:
$$ I = O - NSC \cdot <O_{Flux}>$$
where :

\noindent - $I, O, NSC$ represent respectively, the corrected frame, the
original frame and the flux normalized SC;

\noindent - $<O_{Flux}>$ is the total flux of the original frame.

The SC isn't responsible for the poor accuracy of the sky subtraction with the
flat-field method. A first order computation applied on the $(E_{Sky})_i$
expression demonstrates that, on frames on which the fibers are exposed at
the same level, the SC has no effect.
In fact, we can write:

$$ (E_{Sky})_i = {
{{<SKYspectrum_i> - <NSC_i> \cdot F_S} \over {<FF_i> - <NSC_i> \cdot F_F}}
\over
{{1 \over N} \cdot \sum\nolimits_{j=1}^N {{<SKYspectrum_j> - <NSC_j> \cdot F_S}
 \over {<FF_j> - <NSC_j> \cdot F_F}}}} -1 $$

\noindent where the new symbols $<NSC_i>, F_S, F_F$ represent the normalized
Sky
Concentration flux present in the channel, the total flux of the
astronomical frame and the total flux of the flat-field frame, respectively.
The sum $\sum_j$ is performed over all the $N$ channels exposed on the sky.

Assuming:

$$<SKYspectrum_i> \simeq  \overline{S} \simeq  {F_S \over M}$$
$$<FF_i> \simeq  <\overline{FF}> \simeq {F_F \over M}$$

where $M$ is the total number of the fibers. Then:

$$ (E_{Sky})_i = {{1 - <NSC_i> \cdot ({F_S \over \overline{S}} - {F_F \over
\overline{FF}})} \over {{1 \over N} \cdot \sum\nolimits_{j=1}^N
 (1 - <NSC_j> \cdot ({F_S \over \overline{S}} - {F_F \over
\overline{FF}}))}}-1=0$$

The possible improvement is negligible and the application of the flat-field
method to the SC corrected frames confirms this deduction.
On the contrary,
applying the ``5577'' procedure to the SC corrected frames, the
sky subtraction accuracy improves to $1.6-2.4 \%$ (Table 3).
\medskip

{\it 3.2 The cross-talk influence}

After removing the SC from the mono-fiber flats, we have analyzed the extended
broad tails in the transversal profile of the mono-fiber flats (Figure 3).
We have called transversal profile, or briefly TP, a trace along a
direction perpendicular to the direction of the dispersion
(in our case approximately the same as the y axis).

\vskip 1cm
\centerline{Figure 3}
\vskip 1cm

A first-order argument as in the
case of the SC, shows that the improvement in accuracy on $E_{Sky}$ is
negligible also in this case for frames with fibers exposed at roughly
the same level.

We concluded therefore that the source of unreliability of the ``flat-field''
method has to be ascribed to telecentricity effects and/or non-uniformity of
the flat-field light source.

\medskip

{\bf 4. THE INFLUENCE OF EACH FIBER ON THE ADJACENT ONES}

What are the effects of the cross-talk (in the following Point Spread Function
Background or PSFB) on the ``5577'' method ?

The broadness of the TP obviously indicates that also the PSF of the
instrument has long tails at low flux levels. In principle it is possible
to deconvolve the frame with a given PSF but problems arise (Brault and
White, 1971), due mainly to the type
of sampling and to the boundary conditions required (the signal
doesn't go to zero within the boundaries of the CCD).
We have preferred to produce a simplified method by which we can
subtract the cross-talk effect among the fibers.
We assume that the PSF is a sum of functions of which only the first
component (the main, $PSF_0$) isn't responsible for the cross-talk.
We can write:
$$ PSF(r) = PSF_0(r) + PSF_1(r)$$
The PSF is assumed with a rotational symmetry.
In fact all the emission lines of the helium lamp can be fitted by a two
dimensional gaussian with the same standard deviation ($\sigma_x = \sigma_y
= 1.24 \pm 0.05$ pixels with the dispersion due mainly to differences among
fibers), independent of the wavelength and the position of the channel on the
CCD frame. In this way we can make use of the classical convolution theorem
(Andrews and Hunt, 1977).
The observed TP is the result of the convolution of the spectrum with the PSF:
$$ TP(x) =Spectrum \otimes PSF \|_{y=constant} = TP_0(x) + TP_1(x)$$
where:
$TP_0(x), TP_1(x)$ are due to $PSF_0(r)$ and $PSF_1(r)$ respectively. It is
useful to see that only the study of $PSF_1$ is important because only this
part of the PSF produces the background PSFB. We find that the $TP_1$
can be described by a sum of gaussian functions:

$$TP_1(x) = \sum_k H_k \cdot e^{-{x^{2} \over {2 \cdot {\sigma_k}^2}}}$$

where:

\noindent $H_k, \sigma_k$ are the amplitude and the standard deviations
of the gaussian functions.

If we hypothesize a
$PSF_1$ as sum of gaussian functions, also $TP_1$ is a sum of gaussian
functions (as it is observed) with the same standard deviations
\medtype\footnote{*}
{\rm The conservation of the function type doesn't hold for every function.}
:

$$PSF_1(r) = \sum_k A_k \cdot e^{-{r^{2} \over {2 \cdot {\sigma_k}^2}}}$$

According to the convolution theorem, the PSFB expression in the pixel space
is:

$$F(x_P,y_P) = \int_{x_p-1/2}^{x_p+1/2} \,dt \int_{y_p-1/2}^{y_p+
1/2}\,du \int_{- \infty}^{\infty} PSF_{1}(t,u-y)\,J(y) \cdot \,dy$$

where:

\noindent - $J(y)$ is the flux distribution of the ``true'', ideal spectrum;

\noindent - $x_P, y_P$ are the coordinates of the pixel $P$;

\noindent - $F(x_P,y_P)$ represents the observed flux at the pixel $P$;

\noindent which becomes with a first order development (for $\sigma_k \gg 1$)
and assuming the mono-fiber flat positioned at $x=0$:

$$F(x_P,y_P) = \sum_{k}  A_k \cdot e^{-{{x_P}^2 \over
{2 \cdot {\sigma_k}^2}}} \cdot \sum_{i} j(y_i) \cdot e^{-{(y_P-y_i)^2
\over {2 \cdot {\sigma_k}^2}}}$$

\noindent where $j(y_i)$ is the observed flux of the spectrum at the
pixel coordinate $y_i$;

For a given $y_P=constant$ we obtain a particular $TP_1(x_P)$. The coefficients
$A_k, H_k$ are related by:

$$A_k =  {H_k \over { \sum_{i} j(y_i) \cdot e^{-{(y_P
-y_i)^2 \over {2 \cdot {\sigma_k}^2}}}}}$$

In practice only $3$ gaussian functions are necessary.
In this way from the observed TP at a given $y_p$ it is possible to
estimate by gaussian fitting the $\sigma_k$ and the $H_k$ parameters
(note that the $H_k$ depend on $y_p$) from which the $A_k$ (independent from
$y_p$) are obtained (Table 4).
Then it is possible to estimate the overall background due to the
$PSF_1$ by convolving an artificial frame (in which the flux
contained in each channel has been concentrated in one column corresponding
to the $x$ position of the center of the channel), representing a rough
approximation of a deconvolved image, with the $PSF_1$ (the main
responsible for the cross-talk).
This background PSFB (Figure 4) can now be subtracted from the SC corrected
frame:

$$ II = I - PSFB$$
where :

\noindent - $II, I$ represent respectively, the final frame and the SC
corrected frame.

\vskip 1cm
\centerline{Figure 4}
\vskip 1cm

The improvement of the accuracy (from $1.6-2.4 \%$ to $1.2-1.6\%$, see Table 2)
depends strongly on the distribution of
fluxes among the fibers of the original frame. The ``anomalous''
fiber in the frame $C$ turns out not to be ``anomalous'' any longer.
As shown in Figures 1 and 5 this fiber lies immediately to the right of a
strongly exposed fiber, whose cross-talk was responsible for the ``anomaly''.

\vskip 1cm
\centerline{Figure 5}
\vskip 1cm

Unfortunately the lack of information about the light flux outside the CCD
frame causes a PSFB correction not entirely satisfactory, expecially in the
red part of the spectra.
The variations of the sky subtraction accuracy as a
function af the wavelength range are shown in Table 5.

\vskip 1cm
\centerline{Table 5}
\vskip 1cm

\medskip
{\bf 5. THE EFFECTS OF IMAGE DISTORTIONS ON THE SKY SAMPLING}

With the ``5577'' method some defects of the
``standard'' approach have been eliminated. However, the subtraction of the sky
from the object spectrum leaves embarrassingly large oscillations in the
observed flux in correspondence to strong emission lines and bands. This
problem strongly affects the red part of many spectra making them in practice
unusable beyond 7000 \AA. The usual spectra extraction procedure consists in
the following steps:

- spectra extraction, optimizing the Signal to Noise ratio, e.g. the
procedure EXTRA/LONG in the MIDAS package (Banse et al., 1983);

- calibration in wavelength of the spectra (sky and object spectra), rebinning
with a constant wavelength step;

- mean-sky evaluation;

- sky subtraction;

- absolute flux calibration.

The second step partially removes the image distortions but introduces
secondary effects: not only the rebinning deteriorates the signal to noise
ratio but also small errors in the wavelength calibration of each fiber give
origin to large errors in the subtraction of the narrow sky emission lines. To
overcome this problem we have developed a procedure which tries to map the
image distortions in the pixel space. Using the Seidel distortion theory
(Jenkins and White, 1957):

$$\rho = \rho_0 +E \cdot \rho_0 ^3+ o(5)$$

where:

\noindent - $\rho$ is the distance of a point from the optical axis on
the object plane;

\noindent - $\rho_0$ is the distance of a point from the optical axis
on the image plane;

\noindent - $E$ is the distortion coefficient of the optical system;

\noindent - $o(5)$ represents fifth order terms in $\rho_0$
which are neglected in the third order theory.

In an undistorted coordinate system the expression for a spectrum is:

$$ X=constant$$

\noindent Its expression, in a first-order development, in the distorted
coordinate system (the observed one) is:

$$x = \alpha_i + \beta_i \cdot y + \gamma_i \cdot y^2 $$

\noindent where:

$$
\cases{\alpha_i = Q_i + a - mb \cr
        \beta_i = m - 2b Q_i E     \cr
        \gamma_i = Q_iE }
$$

\noindent where: $Q_i$ are the $x$ positions of the spectra at $y=0$,
$m$ represents the tangent of the angle describing the misalignment of the
$y$ columns of the CCD with respect to dispersion direction, $a$ and $b$
are the $x$ and $y$ coordinates of the optical axis on the CCD frame.

By measuring $\alpha_i,~ \beta_i,~ \gamma_i$ for each spectrum, we can solve
the system and obtain the parameters $a,~ b,~ m,~ E$ (Table 6).

\vskip 1cm
\centerline{Table 6}
\vskip 1cm

The observed $x$ and $y$ coordinates of a given pixel
are mapped on the corrected $Y$ by:

$$Y \simeq y + m x - E \cdot (y-b) \cdot
 [(x-a)^2+(y-b)^2]$$

\noindent where $m x$ represents the effect of the rotation
(typically $< 0.1$ pixel, see Table 6)
and $E \cdot (y-b) \cdot [(x-a)^2+(y-b)^2]$
the effect of distortion (it can easily exceed 1 pixel).
Only the $Y$ coordinate is actually relevant because it represents the position
along the dispersion direction of the resulting spectrum.
After this correction the extracted spectra (monodimensional) show
different dispersion scales
and zero points, which can be estimated using the wavelength calibration
exposures.
The behaviour of the dispersion scales is shown in Figure 6.

\vskip 1cm
\centerline{Figure 6}
\vskip 1cm

To bring all the spectra to the same wavelength scale and zero point it
is necessary to introduce a further correction:

$$ Yc_{i} = coef_i \cdot Y + q_i $$

\noindent where:

\noindent $Yc_{i}$ indicates the corrected coordinate
for the spectrum $i$ ;

\noindent $coef_i, ~q_i$ are the dispersion scales and the zeropoint shift
values of the $i$ spectrum, estimated by comparing the positions of
He-Ar emission lines with a reference spectrum.
{}From Figure 6 it is possible to deduce that the effects due to the different
scales imply displacements of the order of half a pixel.

To determine the residual shifts, essentially due to differences between the
optical path of astronomical observations and He-Ar calibrations
(Figure 7, bottom panel), we can use
a few emission sky lines (e.g. the [OI] 5577 \AA, 6300 \AA) and carry out
the final correction:

$$ Ye_{i} = Yc_{i} + r_i$$

\noindent where:

\noindent $Ye_{i}$ are the final corrected coordinates
for the spectrum $i^{th}$;

\noindent $r_i$ is the shift of the $i^{th}$ spectrum.

After all these corrections the sampling of each spectrum is different from
the other ones.
The union of all the sky normalized spectra (the $SKY_i$ spectra) allows
the building of an ``oversampled'' global sky spectrum.
By interpolation and rebinning of this ``oversampled'' spectrum we obtain an
estimate of the sky in the reference system of wavelengths of each
object spectrum.

\vskip 1cm
\centerline{Figure 7}
\vskip 1cm

The advantages of this sky subtraction technique are:

- the drastic reduction of the oscillations in the sky subtracted spectra
in correspondence of narrow emission features

- preservation of the true signal to noise ratio and of the
independence of the flux of each pixel, allowing also a correct estimate
of the S/N

Typical results are shown in Figure 8 and 9.

\vskip 1cm
\centerline{Figure 8 and 9}
\vskip 1cm

The procedures described in sections 4 and 5 are illustrated in the flow
chart of Figure 10, representing a set of procedures developed in the
framework of the MIDAS image processing system (Banse et al, $1983$).

\vskip 1cm
\centerline{Figure 10}
\vskip 1cm

\medskip

 {\bf 6. CONCLUSIONS}

The poor and unreliable sky subtraction often observed when operating
a multifiber spectrograph depends strongly on the method used to
estimate the transmittance of the fibers.
A method based on the observed flux of the 5577 [OI] line
has been shown to allow a considerable increase in the accuracy of the
sky subtraction.

Further improvements are obtained when the effects of the internal scattered
light and the influence of each fiber on the adjacent ones are
determined and corrected.

For the OPTOPUS spectrograph, this allows to reach a sky subtraction
accuracy of $1.2$-$1.6 \%$.

To obtain the maximum performance from the instrument it is necessary to
develop optimized extraction and subtraction procedures that take into
account also the effects of the image distortions on the sky sampling.

The overall improvement is important in view of the application of multifiber
instruments on telescopes of $6$-$10~meters$ diameter (VLT, KECK, etc..).
With low dispersion ($ \sim $ 10 \AA/pix )
and reasonably long exposures ($ \ge \sim $ 1 hour) the main limitation to the
S/N already for telescopes of the 4m class is due to the sky subtraction,
if its accuracy is limited to the $5$-$8 \%$ range. In such a situation
it would not be useful in terms of limiting flux to apply multi-fiber
spectrographs on bigger telescopes. The techniques described in this paper
demonstrate that this limit can be significantly pushed down, making worthwhile
the construction of multi-fiber spectrographs for very large telescopes.
\bigskip

{\bf ACKNOWLEDGEMENTS:} We thank P. Andreani, G.Avila and S. D'Odorico
for enlightening discussions and suggestions.

\vfill
\eject

\centerline {\bf REFERENCES}

\ref{Andrews, H.C. and  Hunt, B.R., 1977 {\it Digital Image Restoration},
Prentice-Hall Signal Series.

\ref{Avila, G. and D'Odorico, S., 1993 {\it Fiber Optic in Astronomy}, ASPC.
{\bf 37}, 90}.

\ref{Banse, K., Crane, P., Ounnas, C. \& Ponz, D., 1983. In: {\it Proc. of
DECUS}, Zurich, p. 87.}

\ref{Brault, J.W., White, O.R., 1971, A\&A, {\bf13}, 169.

\ref{Elston, R. and Barden, S., 1989, NOAO {\it Newsletter}, 9/89, p. 21.

\ref{Jenkins, F.A., White, H.E., 1957, {\it Fundamentals of Optics},
ed. McGraw-Hill, New York, pg. 132}

\ref{Mignoli, M. and Cuby, J.G., 1994, Proc. of SPIE conference {\bf 2198},
in press.}

\ref{Schwarz, H.E. and Melnick, J., 1993 {\it ESO Users Manual}}.

\ref{Wynne, Ch.G., 1993, MNRAS,
{\bf 260}}, 307.

\ref{Wyse, R.F.G. and Gilmore, G., 1992, MNRAS,
{\bf 257}}, 1.

\vfill
\eject


\centerline {\bf FIGURE CAPTIONS}
\bigskip

{\bf Figure 1:} A typical raw CCD frame obtained with OPTOPUS (in this case
frame C). Notice the 22$^{nd}$ fiber from the left, strongly exposed
at the ambient light. The ``anomalous'' fiber discussed in the text
lies immediately next to the right.
\bigskip

{\bf Figure 2:} Representation of the Sky Concentration (in units of
$10^{-6}$).
\bigskip

{\bf Figure 3:} The mean Transversal Profile (see text).
\bigskip

{\bf Figure 4:} The Point Spread Function Background (PSFB) for the frames
D and A. Notice the effect of the one strongly exposed fiber in frame D.
\bigskip

{\bf Figure 5:} The calculated PSFB at row $300$ of frame C.
\bigskip

{\bf Figure 6:} The variation of the dispersion coefficient versus the
coordinate of each channel.
\bigskip

{\bf Figure 7:} Effect of the residual shifts $q_i$ between the calibration and
object frames (frame B). Top: all the sky spectra in the $Yc_i$ system
coordinate (see text). Middle: the same spectra in the $Ye_i$ system
coordinate. Bottom: the shift
values versus  the center coordinate of the channel along the x axis.
\bigskip

{\bf Figure 8:} Spectra of the same object extracted
with the ``standard'' reduction (top) and with the procedure described
in the text (bottom).
\bigskip

{\bf Figure 9:} A new quasar (continuous line)
discovered in the field of SA94, compared with the sky flux
(dashed line): only with the new
procedure it has been possible to extract and identify the characteristic
emission lines of this spectrum.
\bigskip

{\bf Figure 10:} Flow chart of a reduction session of OPTOPUS frames using the
procedures produced within the framework of the MIDAS image processing system.
\bigskip

\bye